\documentclass[12pt]{iopart}
\usepackage[T1]{fontenc}
\usepackage[ansinew]{inputenc}

\begin{document}
%\pagecolor{black}
%\color{white}

\title[General Relativity as Classical Limit
of Evolutionary Quantum Gravity]{General Relativity as Classical Limit
of Evolutionary Quantum Gravity}

\author{Giovanni Montani$^{123}$, Francesco Cianfrani$^{1}$}

\address{$^{1}$ICRA---International Center for Relativistic Astrophysics\\ 
Dipartimento di Fisica (G9),\\ 
Universit\`a  di Roma, ``La Sapienza",\\ 
Piazzale Aldo Moro 5, 00185 Rome, Italy.\\ 
$^{2}$ENEA C.R. Frascati (Dipartimento F.P.N.),\\
Via Enrico Fermi 45, 00044 Frascati, Rome, Italy.\\
$^{3}$ICRANet C. C. Pescara, Piazzale della Repubblica, 10, 65100 Pescara, Italy.}

\ead{montani@icra.it\\
francesco.cianfrani@icra.it}

\begin{abstract}

We analyze the dynamics of the gravitational field when the covariance is restricted to a synchronous gauge. In the spirit of the Noether theorem, we determine the conservation law associated to the Lagrangian invariance and we outline that a non-vanishing behavior of the Hamiltonian comes out. We then interpret such resulting non-zero ``energy'' of the gravitational field in terms of a dust fluid. This new matter contribution is co-moving to the slicing and it accounts for the ``materialization'' of a synchronous reference from the corresponding gauge condition. Further, we analyze the quantum dynamics of a generic inhomogeneous Universe as described by this evolutionary scheme, asymptotically to the singularity. We show how the phenomenology of such a model overlaps the corresponding Wheeler-DeWitt picture. Finally, we study the possibility of a Schr\"odinger dynamics of the gravitational field as a consequence of the correspondence inferred between the ensemble dynamics of stochastic systems and the WKB limit of their quantum evolution. We demonstrate that the time dependence of the ensemble distribution is associated with the first order correction in $\hbar$ to the WKB expansion of the energy spectrum.
\end{abstract}

\pacs{83.C}

\maketitle

\section{INTRODUCTION}

The absence of a real time evolution of the
physical states for the quantum gravitational field, is
one of the most peculiar aspects characterizing the
Wheeler-DeWitt equation \cite{D67}.
It emerges as a direct consequence of implementing on
a quantum level the 4-diffeomorphisms invariance of
General Relativity. In fact, in the sliced picture of
the space-time, the manifold $\mathcal{V}^4$ is
represented by a one-parameter family of spacelike
hypersurfaces ({\it i.e.}
$\mathcal{V}^4\rightarrow \sigma ^3_t\otimes \mathcal{R}$)
and the dynamics is summarized by the primary and
secondary constraints, due to the presence of
four Lagrangian multipliers (the lapse function and
the shift vector) \cite{ADM62}. Hence, extending the
canonical Dirac methods of quantization constraints to
the gravitational sector, the
\emph{frozen formalism} arises \cite{K81}.
For a detailed discussion of the problem of time
in quantum gravity and for a review of different
proposals to overcome it, see \cite{I92}
(about the nature of time in quantum cosmology,
see \cite{SS04}, while for an evolutionary scenario coming out in the semi-classical limit see \cite{Ba85}). A valid discussion of the relation
existing among time, matter, and reference frames in
canonical quantum gravity is given in \cite{Ro91a,Ro91b}.

In \cite{M02} and \cite{MM04} it was inferred that
the non-evolutionary character of the Wheeler-DeWitt
equation is a consequence of requiring that the 3+1-splitting of
the space-time holds also on a quantum level.
The point is that, in a covariant picture,
the canonical quantization applies only if a physical
reference fluid is included into the dynamics.
In fact, the timelike character
of the 4-velocity associated to a fluid has to be
preserved in a quantum space-time too
and it allows a physical slicing.
The analysis presented in
\cite{M02,MM04} includes the so-called kinematical
action into the evolution and shows
how the resulting
``frame fixing'' quantization
of the vacuum gravitational field
induces the appearance
of a matter fluid as a source. 
The approach based on the kinematical action
can be re-casted as a Schr\"odinger dynamics for
the quantum gravitational field \cite{MM04b}.
This same point of view was also
addressed in \cite{KT91}, where
it is outlined how the quantum
gravitational field, viewed in a
synchronous (or Gaussian) frame,
acquires an evolutionary character and a dust fluid
arises into the dynamics
(see also the related discussion in \cite{BK95}).
Other important approaches based on the
so-called 
embedding variables, and even referred to
the path integral formalism, can be found
in \cite{H91a}-\cite{H91c}
(see also \cite{Tom}).\\

Here  we face the classical and the quantum dynamics of
the synchronous gravitational field, starting from a
restriction of the covariance principle to those coordinates
transformations which preserve the choice of this gauge.
The phenomenological issue of the synchronous quantum gravity, 
so defined, outlines the appearance of a non-vanishing eigenvalue
Hamiltonian, reflecting the presence of a dust fluid. 
Since the privileged role of a dust fluid as a physical clock is well-established in
literature, we have to point out the peculiar aspects
of our approach. In \cite{KT91}, \cite{BK95}
and \cite{M02} the construction of the clock is based
on adding new terms to the system action, and then deriving the new
Hamiltonian constraints. Here we focus
attention on the symmetry of the synchronous space-time,
having in mind that the choice of a coordinate system
must come out into additional energy-momentum contributions.
We put in correlation the (restricted) symmetry invariance
with the quantum nature of the appearing fluid.
In fact, the violation of the general relativity
principle singles out by the appearance of a source term,
living in a covariant picture and whose energy
(expected to be positive in the ground state) vanishes in
the classical limit.
The main new address of our investigations 
is essentially in this idea, that General Relativity is
compatible with the synchronous quantum gravity.
The discussion of Section 6 clarifies this point
of view, by underlining the link between the restored
quantum time variable and the
spectrum dependence on $\hbar$.\\ 
This new contribution can be heuristically interpreted as the
quantum ``materialization'' of the synchronous gauge imposed
on the vacuum theory of gravity. We also address the request of having a positive energy density of the dust
and we infer that the ground state of the theory ensures such requirement is fulfilled. The main point is that the magnitude of the eigenvalue arising from the generic quantum cosmology is bounded, of the order of $\hbar$, and vanishing in the classical limit $\hbar \rightarrow 0$.
Hence, we discuss the possibility of a general character
for this feature, and we provide an implementation of this
point of view within the correspondence existing, for chaotic systems,
between the ensemble distribution and the semiclassical
wave function.

Thus, we conclude that a scenario can be inferred in which
the time evolution of the quantum gravitational
field takes place only at a higher order in the $\hbar$ expansion
of the theory. In this respect, the phenomenology of this
evolutionary quantum gravity overlaps the same issues
of the Wheeler-DeWitt approach, and General Relativity is
recovered in the classical limit. This result is a consequence of the ``quantum'' character of the device responsible for the emergence of time, thus providing an explanation for the applicability of the approach discussed in \cite{PW83} into a cosmological setting. By other words, we fix a time-clock which turns out to be a quantum component of the whole system and so the notions of external and internal times converge.\\

This paper is organized as follows. In Section 2,
we derive the fundamental constraints implied on the theory by the invariance of the Lagrangian, in the
framework of a Noether theorem extended to the gravitational sector.
Section 3 is devoted to discuss the canonical quantization
of the synchronous gravitational field, and the
question concerning the physical interpretation of the
outcoming Hamiltonian eigenvalue.
In Section 4, we formulate the cosmological problem inherent to
a generic inhomogeneous Universe in the presence of a
massless scalar field and of a cosmological term, which allow to
model an inflationary scenario. We develop the canonical quantization
of this model in the framework of
a Schr\"odinger dynamics. The possibility to
neglect the potential term, in the asymptotic limit to
the cosmological singularity, allows to deal with
an approximated analytic solution. The precise conditions
for the validity of the proposed picture are as the ones
for the existence of an inflationary scenario.
The Hamiltonian eigenvalue comes out as ranging,
in modulus, between zero and much less than the
Planck energy. Since a negative portion of the
spectrum arises, an estimate for the contribution
of dust to the Universe critical parameter is
given (assuming the Universe near its ground state).
Such a contribution is extremely small
$\mathcal{O}\left(10^{-60}\right)$ and therefore
we are lead to phenomenologically recover no
observability for a primordial quantum evolution
of the Universe.\\
In Section 5, we discuss the quasi-classical limit of the model,
which outlines how the variable associated to the Universe
volume reaches the classical stage before the potential term
becomes relevant in the dynamics.

In Section 6, we provide a discussion concerning the formulation
of an evolutionary quantum gravity from a more general and
gauge-independent context. The approach is based on comparing
the ensemble representation of a stochastic system with 
the semiclassical WKB limit of its quantum dynamics. In particular, Section 6.1 is devoted to fix the paradigm of
such a correspondence for a generic stochastic gravitational field. Here we show that
a non-zero super-Hamiltonian eigenvalue is expected to reproduce
the right behavior of the ensemble distribution. Section 6.2 discusses the implementation of the outlined scheme
to the case of the inhomogeneous mixmaster model as
a gravitational stochastic system.

Finally, in Section 7 we give some concluding remarks about
the main lines of thinking fixed by the overall analysis.

\section{GRAVITY IN A SYNCHRONOUS REFERENCE}\label{2}

In a synchronous (gaussian) reference frame to the splitting $y^\mu=y^\mu(t,x^i)$, the metric tensor corresponds to the
choice $g_{00} = 1$ and $g_{0i} = 0$
($i=1,2,3$), {\it i.e.} in the 3+1-formalism
we have to require $N = 1$ and $N^i = 0$ for the lapse function and for the shift vector, respectively. In order to fix the form of the coordinates transformations
which preserve the synchronous character
\cite{KL63},
we consider a generic infinitesimal displacement

\begin{equation}
t^{\prime } = t + \xi(t,\; x^l)\,
\quad x^{i^{\prime }} = x^i + \xi ^i(t,\; x^l)
\label{srp1}
\end{equation}

and the associated 4-metric change

\begin{equation}
g^{\prime }_{\mu \nu} = g_{\mu \nu } - 2~^4\nabla _{(\mu }\xi _{\nu )}
\, ,
\label{srp2}
\end{equation}

with $\xi ^{\mu } = \{\xi,\; \xi^i\}$
($\mu = 0,1,2,3$). 
From (\ref{srp2}),
preserving $g_{00} = 1$
and $g_{0i} = 0$, it comes out that the following
two conditions respectively hold

\begin{eqnarray}
\label{srp3}
\partial _t\xi = 0 \; \Rightarrow \;
t^{\prime } = t + \xi (x^l)\\
h_{ij}\partial _t\xi^J = \partial _i\xi \; \Rightarrow \;
x^{i^{\prime }} = x^i
+ \partial _j\xi \int h^{ij}dt + \phi ^i(x^l)
\, , \label{srp3bis}
\end{eqnarray}

where $\phi^i$ denote three generic space functions.
Finally, for the 3-metric we get the transformation

\begin{equation}
h^{\prime }_{ij}(t^{\prime },\; x^{l^{\prime }}) =
h_{ij}(t^{\prime },\; x^{l^{\prime }})
- 2~^3\nabla _{(i}\xi _{j)} -
\partial _th_{ij}\xi 
\, .
\label{srp4}
\end{equation}

In a synchronous reference, 
the Lagrangian of the gravitational field, in presence of a cosmological constant $\Lambda$, reads

\begin{equation}
L_{grav} = \int _{\Sigma ^3_t}d^3x \mathcal{L}_{grav} = 
-\frac{1}{2c^2k}\int _{\Sigma ^3_t}
d^3x\sqrt{h} \left\{ K^2 - K_{ij}K^{ij} - ~^3R\right\} -\frac{1}{k}\int _{\Sigma ^3_t}
d^3x\sqrt{h}\Lambda 
\, ,
\label{srp5}
\end{equation}

where $k$ is the Einstein constant
($k = 8\pi G/c^4$), $h\equiv det h_{ij}$ and
$K_{ij} \equiv -\frac{1}{2}\partial _th_{ij}$
refers to the extrinsic curvature, while $K \equiv h^{ij}K_{ij}$. 
In terms of the Lagrangian density (of weight $1/2$)
$\mathcal{L}$, the $i-j$ components of the
Einstein equations take the Euler-Lagrange form

\begin{equation}
\partial _t
\left( \frac{\delta L}{\delta (\partial_th_{ij})}
\right) + 
\partial _l
\left( \frac{\delta L}{\delta (\partial_lh_{ij})}
\right) - 
\frac{\delta L}{\delta h_{ij}} = 0
\, .
\label{srp6}
\end{equation}

Under an infinitesimal 3-metric displacement
$h^{\prime }_{ij}(t^{\prime },\; x^{l^{\prime }})
- h_{ij}(t,\; x^l) =
\delta h_{ij} + \partial _th_{ij}\xi$, 
the Lagrangian  density $\mathcal{L}$ changes correspondingly as
(where, the contribution $\delta h_{ij}$ is provided by (\ref{srp4})
and the 3-metric is transported parallel
along the space hypersurfaces)

\begin{eqnarray}
\label{srp8}
\delta \mathcal{L}
= \mathcal{L}_{grav}\left( h^{\prime }(x^{\prime }), \; \partial ^{\prime }
h^{\prime }(x^{\prime })\right) - \mathcal{L}_{grav}\left( h(x), \; \partial h(x)\right)+\delta \mathcal{L}_{mat} =\\\nonumber 
= \mathcal{L}_{grav}\left( h^{\prime }(x^{\prime }), \; \partial ^{\prime } h^{\prime }(x^{\prime })\right) - 
\mathcal{L}_{grav}\left( h(x^{\prime }), \; \partial ^{\prime }h(x^{\prime })\right) +
\partial _t\mathcal{L}_{grav}\xi+\partial _i\mathcal{L}_{grav}\xi^i+\delta \mathcal{L}_{mat},
\end{eqnarray}

where we adopted a schematic notation for the sake of simplicity.\\
Since in the following we will see the application to a cosmological setting with a scalar field, we are going to consider the case in which a scalar field $\phi$ is present.\\ 
Hence, let us consider the full Lagrangian density $\mathcal{L}=\mathcal{L}_{grav}+\mathcal{L}_{\Phi}$. The analogous of the expression (\ref{srp4}) for the adopted matter field is the following one 
\begin{equation}
\varphi'(x^{l'},t')=\varphi(x^{l'},t')-\xi^i\partial_i\varphi-\xi\partial_t\varphi\qquad ,\label{mat}
\end{equation}
while Euler-Lagrange equations are obtained from the ones for the gravitation field, by replacing $h_{ij}$ with $\varphi$.\\
Thus, the invariance request explicitly reads

\begin{eqnarray}
\delta L =
\frac{\delta L}{\delta h_{ij}}\delta h_{ij} + 
\frac{\delta L}{\delta (\partial _l h_{ij})}
\delta (\partial _l h_{ij}) + \frac{\delta L}{\delta (\partial _t h_{ij})}
\delta (\partial _t h_{ij}) +\nonumber\\ +\frac{\delta L}{\delta\varphi}\delta \varphi + 
\frac{\delta L}{\delta (\partial _l\varphi)}
\delta (\partial _l\varphi)+ \frac{\delta L}{\delta (\partial _t\varphi)}
\delta (\partial _t\varphi)
+ \int \partial _t \mathcal{L}\xi d^3x = 0 
\, ,
\label{srp7i}
\end{eqnarray}
where a 3-divergence has been eliminated by suitable conditions at spatial boundary.
Making use of equations (\ref{srp6}) and avoiding other 3-divergences, 
we finally arrive to the conservation law

\begin{equation}
\partial _t\left\{ \int _{\Sigma ^3_t}d^3x\left[ 
\frac{\delta \mathcal{L}}{\delta (\partial _th_{ij})}
\delta h_{ij} + \frac{\delta \mathcal{L}}{\delta (\partial _t\varphi)}
\delta \varphi + \mathcal{L}\xi \right] \right\} = 0 
\, . 
\label{srp11}
\end{equation}

Substituting (\ref{srp4}) and (\ref{mat}) in the above relation,
and observing that
$\frac{\delta \mathcal{L}}{\delta (\partial _th_{ij})}$ and $\frac{\delta \mathcal{L}}{\delta (\partial _t\varphi)}$ 
give the conjugate momenta
$\pi ^{ij}$ and $\pi$ to the variables $h_{ij}$ and $\varphi$ respectively, 
we rewrite (\ref{srp11}) in the form

\begin{equation}
\partial _t\left\{ \int _{\Sigma ^3_t}d^3x\left[ 
-2\pi ^{ij} ~^3\nabla _i\xi _j- \xi^i\pi\partial_i\varphi
-\left( \pi^{ij}\partial _th_{ij}+\pi\partial _t\varphi
-\mathcal{L}\right) \xi \right] \right\} = 0
\, . 
\label{srp12}
\end{equation}

Above, the second term in parentheses coincides with the
super-Hamiltonian $H$, while the first one, by virtue of (\ref{srp3}) and (\ref{srp3bis})
and integrating by parts, can be restated as

\begin{equation}
\int _{\Sigma ^3_t}d^3x\left[
(2~^3\nabla _j\pi _i^j-\pi\partial_i\varphi)\left( \phi ^i +
\partial _l \xi \int dth^{il}\right)
\right] 
\, . 
\label{srp13}
\end{equation}

Recalling that the super-momentum $H_i$ is given by
$-2~^3\nabla _j\pi _i^j+\pi\partial_i\varphi$, taking into (\ref{srp12}) the time derivative and using the
relation $\partial _t\xi ^i =  \partial _l\xi h^{li}$, 
we see that the invariance request reads

\begin{equation}
\int _{\Sigma ^3_t}d^3x\left\{ 
-\partial _t(H_i)\left( \phi^i + \partial _l\xi \int dt h^{il}
\right) - \left( \partial _tH - \partial _lH^l\right) \xi \right\} = 0
\, . 
\label{srp14}
\end{equation}

Since $\xi^{\mu } = \{ \xi ,\; \xi ^i\}$ are four generic
(independent) displacements, then the solution
to the above integral equation is provided
by the following constraints

\begin{equation}
\partial _tH_i = 0 \; , \; \partial _tH = \partial _lH^l 
\, .
\label{srp15}
\end{equation}

The first three constraints yield $H_i = k_i(x^l)$, reducing
the fourth one to $\partial _tH = \partial _lk^l$. We now observe
that $H$ does not depend explicitly on time while, 
on the other hand, the super-momentum constraints have to remain
independent by each other because the pure 3-diffeomorphisms are included into
the transformations (\ref{srp3}) in correspondence
to $\xi \equiv 0$ ({\it i.e.} we must have $\partial _ik^i \neq 0$,
if $k_i \neq 0$).
Therefore, the only available solution to the system (\ref{srp15})
stands as the following constraints 

\begin{equation}
H^* \equiv H - \mathcal{E}(x^l) = 0 \, \quad H_i = 0 
\, . 
\label{srp16}
\end{equation}

The obtained result outlines how, preserving 
in geometrodynamics the
synchronous character of the reference, we are lead to
a non-vanishing super-Hamiltonian, while the 3-diffeomorphisms
invariance still holds due to the constraint $H_i = 0$.\\
Since $\mathcal{E}$ is a scalar density
of weight $1/2$, then we can take it in the
form $\mathcal{E}\equiv -2\sqrt{h}\rho (t,\; x^i)$
($\rho$ being a scalar space-time function). 
The action associated to this system of constraints corresponds to the following modification of the Einstein-Hilbert one, in presence of a cosmological constant and of a scalar field,
\begin{equation}
S=-\frac{1}{2c^2k}\int d^4x\sqrt{-g}(R-2\rho\frac{(N-1)}{N})-\frac{1}{k}\int d^4x\sqrt{-g}\Lambda+\int d^4x\sqrt{-g}\mathcal{L}_\varphi,
\end{equation}
$\rho$ being a Lagrangian multipliers, which preserves the co-moving character of the reference. Hence the Hamiltonian is given by 
\begin{equation}
\mathcal{H}=\int d^3x (N(H-\mathcal{E})+N^iH_i)
\end{equation}
and, once a canonical symplectic structure is introduced, the algebra of constraints is as follows
\begin{eqnarray}
\{H_i(x),H_j(y)\}=H_i(x)\delta_{,j}(x-y)-H_j(y)\delta_{,i}(x-y)\\
\{H^*(x),H_i(y)\}=-H^*(y)\delta_{,i}(x-y)\\
\{H^*(x),H^*(y)\}=H^i(x)\delta_{,i}(x-y)-H^i(y)\delta_{,i}(x-y).
\end{eqnarray}
We now observe that the constraints $\{ H^*=0,\: H_i=0\}$
still obey a closed algebra for their Poisson brackets
(see \cite{K81} and \cite{MM04b}).

By its role in the super-Hamiltonian constraint, 
the function $\rho$ acquires the physical meaning
of energy density associated to a co-moving dust fluid.\\

In fact, we are describing a scenario in which a source for the Einstein equation is present, which is co-moving with the slicing and provides the modification (\ref{srp16}) to the constraints of General Relativity. In presence of a fluid with an equation of state $p=(\Xi - 1)\rho$, the Einstein system in the slicing picture reads
\begin{equation}
\rho = -\frac{H}{2\sqrt{h}}, \qquad H_i = 0, \qquad G_{\mu \nu }\partial _iy^{\mu }\partial _jy^{\nu }
\equiv G_{ij} = \kappa (\Xi -1)\rho h_{ij}
\, .
\end{equation}

Hence, the conservation law for the energy-momentum tensor, {\it {\it i.e.}} 
$T_{\mu ;\nu}^{\nu } = 0$
implies the following two conditions

\begin{eqnarray}
\Xi \left( \rho u^{\mu }\right) _{;\mu } =
(\Xi - 1) u^{\mu }\partial _{\mu }\rho \label{geo}\\ 
u^{\nu } u_{\mu ;\nu } = 
\left( 1 - \frac{1}{\Xi }\right)
\left( \partial _{\mu } \ln \rho - u_{\mu }u^{\nu }
\partial _{\nu }\ln \rho \right)
\, .\label{geo2}
\end{eqnarray}

Once the splitting is adapted to the fluid, which means setting the vector normal to the splitting
$n^{\mu } \equiv u^{\mu } = \delta _0^{\mu }$, a Gaussian geodesics frame is fixed ($N=1$ and $N^i=0$) and the consistency of the equation (\ref{geo2}) requires $\Xi = 1$ (dust fluid). In this case, a solution of equation (\ref{geo}) is given by
\begin{equation}
\left( \rho u^{\mu }\right) _{;\mu } =\frac{1}{\sqrt{-g}}\left(\sqrt{-g}\rho u^{\mu }\right) _{,\mu }=\sqrt{h}\partial_t\rho=0\rightarrow \rho = -\bar{\epsilon }(x^l)/2\sqrt{h}.
\end{equation}

The most natural way of thinking about the appearance of
such a new contribution is that the reference fixing procedure
requires a physical realization of the synchronous gauge. 

Furthermore, the allowance for such a new source contribution preserves the 4-diffeomorphism invariance, even though fixing a synchronous reference frame leads to modified Hamiltonian constraints.

Here two main points call for attention.\\
i)---The energy density $\rho$ is not always positive.\\
ii)---The quantity $\mathcal{E}(x^i)$ is fixed by the
initial conditions we assign on a non-singular hypersurface
and therefore it can be, in principle, fixed as arbitrarily
small.

\section{CANONICAL QUANTIZATION OF THE MODEL}

By the Lagrangian 
(\ref{srp5}), the Hamiltonian density
({\it i.e.} the super-Hamiltonian in a synchronous reference frame)
takes the explicit form

\begin{equation}   
H \equiv 2c^2kG_{ijkl}\pi^{ij}\pi^{kl}
- \frac{1}{2k}\sqrt{h}{}^3R  \,, \quad
G_{ijkl} \equiv \frac{1}{2\sqrt{h}}
(h_{ik}h_{jl} + h_{il}h_{jk} - h_{ij}h_{k})\\ 
\, ,
\label{synham}
\end{equation}

where ${}^3R$ denotes the 3-dimensional Ricci scalar.

The canonical quantization of the synchronous gravitational
field is achieved by upgrading the canonical variables
$h_{ij}$ and $\pi ^{ij}$ to operators acting on the state
function $\chi $, {\it i.e.} 

\begin{eqnarray}
\label{qo}
h_{ij}\rightarrow \hat{h}_{ij}\qquad
\pi ^{ij}\rightarrow \hat{\pi }^{ij} = 
-\frac{i\hbar}{(2ck)^{3/2}}
\frac{\delta (\quad )}{\delta h_{ij}}
\end{eqnarray}

and then implementing the synchronous constraints
$H^* = 0$ and $H_i = 0$ as follows

\begin{eqnarray}
\label{nhc}
{\hat{H}}^*\chi _{\mathcal{E}} = 0 
\, \quad \Rightarrow \, \quad 
\hat{H}\chi _{\mathcal{E}} =
\mathcal{E}\chi _{\mathcal{E}} \\           
{\hat{H}}_i\chi _{\mathcal{E}} = 0 
\, .
\end{eqnarray}

To safe the Hermitianity of the 
super-Hamiltonian, we are lead to
take the operator ordering (see \cite{M02}) 

\begin{equation}
G_{ijkl}\pi ^{ij}\pi ^{kl} \rightarrow
\hat{\pi }^{ij}G_{ijkl}\hat{\pi }^{kl}
\, . 
\label{ag11}
\end{equation}

Being the super-Hamiltonian non-vanishing, it turns out that the dynamics is fixed 
by the Schr\"odinger equation 

\begin{equation}
i\hbar \partial _t\chi =
\int _{\Sigma ^3_t}\hat{H} d^3x  \chi 
\, ,
\end{equation}

and the wave-functional evolves with the label time.

The interpretation of the super-Hamiltonian eigenvalue
as physical matter, relies on the proof that
a region of positive energy density exists.

Having in mind this idea, we
adopt more convenient variables to express the
3-metric tensor, {\it i.e.}

\begin{equation}
h_{ij} \equiv \eta ^{4/3}u_{ij}
\, ,
\label{detvar}
\end{equation}

with $\eta \equiv h^{1/4}$ and $det u_{ij} = 1$.\\
Expressed via these variables, the synchronous action reads

\begin{equation}   
S = \int _{\Sigma ^3_t} \left\{
p _{\eta }\partial _t\eta +
p ^{ij}\partial _tu_{ij}
- H\right\} d^3xdt
\, ,
\label{sac}
\end{equation}  

where $p_{\eta }$ and $p^{ij}$ denote the conjugate momenta to
$\eta $ and $u_{ij}$ respectively, while the Hamiltonian density
takes the form

\begin{equation}
H = -\frac{3}{16}c^2 kp_{\eta}^2+
\frac{2c^2 k}{\eta ^2}u_{ik}u_{jl}p^{ij}p^{kl}
- \frac{1}{2k}\eta ^{2/3}V(u_{ij},\; \nabla \eta ,\; \nabla u_{ij})
\label{newh}
\, .
\end{equation}

Here, the potential term $V$ comes from the 3-Ricci scalar and
$\nabla$ refers to first and second order spatial gradients.

In this picture, 
the first of the equations (\ref{nhc})
takes the form

\begin{equation}
\label{newhaux}
\hat{H}\chi _{\mathcal{E}} =
\left\{ 
\frac{3}{128\hbar ck^2}\frac{\delta ^2}{\delta \eta ^2} -
\frac{1}{4\hbar ck^2\eta ^2}\Delta _{u}                          
- \frac{1}{2k}\eta ^{2/3}V(u_{ij},\; \nabla \eta ,\; \nabla u_{ij})
\right\}\chi _{\mathcal{E}}  = 
\mathcal{E}\chi _{\mathcal{E}}
\end{equation}
\begin{equation}
\Delta _{u} \equiv
\frac{\delta }{\delta u_{ij}}
u_{ik}u_{jl}
\frac{\delta }{\delta u_{kl}}
\, ,
\end{equation}

From a qualitative point of view, the existence of solutions
for the system (\ref{newhaux}) with negative values of
$\mathcal{E}$ can be inferred from its Klein-Gordon-like
structure. However, the Landau-Raichoudhuri theorem states
that, in a synchronous reference, the metric determinant always
vanishes monotonically in correspondence to an instant of time
$t^*$ where all the geodesics lines cross each other, i.e
$\eta (t^*, x^i) = 0$, with $\partial _{t\rightarrow t^*}\eta > 0$. 
Such a classical property of the variable $\eta $,
on one hand supports its meaning of internal time and,
on the other one, it allows us to take the limit
$\eta  \rightarrow 0$, where the system 
(\ref{newhaux}) admits an asymptotic solution.
In fact, in this limit, the potential term is
drastically suppressed with respect to the
$\Delta _{u}$ one
and the dynamics of different spatial points decouples.
Thus, the quantization scheme reduces to a
local minisuperspace approach.\\
It is easy to see that such approximate
dynamics admits, point by point in space, the solution

\begin{equation}
\chi _{\mathcal{E}} =
\iota _{\mathcal{E}}(\eta , p)G_{p^2}(u_{ij})
\, ,
\label{iog}
\end{equation}

$\iota $ and $G_{p^2}$ satisfying respectively the two equations

\begin{eqnarray}
\label{twoeig1}
\left\{ 
\frac{1}{\hbar ck^2}\frac{\delta ^2}{\delta \eta ^2} +
\frac{32p^2}{\hbar ck^2\eta ^2}                                    
\right\}\iota _{\mathcal{E}} = 
\mathcal{E}\iota _{\mathcal{E}}\\
\Delta _{u} G_{p^2} = -p^2G_{p^2}
\, .
\end{eqnarray}

%%%%%%%%%%%%%%%%%%%%%%%%%%%%%%
The potential term is negligible, also on a quantum level, as soon as the following condition holds
\begin{equation}
\frac{p^2}{4\hbar ck^2\eta ^2}\gg \frac{1}{2k}\eta ^{2/3}\frac{1}{\Delta u}\int_{\Delta u} d^5u V.
\end{equation}
This relation stands for a wide range of $p^2$ values, approaching $\eta=0$, and it can be obtained by considering a wave packet laying over a region $\Delta u\sim1/\Delta p\gg 1$, where $\Delta p$ is a small uncertainty around the picked value $p$ ($p\gg\Delta p$). Hence the last condition singles out values of $p$ greater than a fiducial one $p_0\sim\Delta p$, according to the prescription that a quantum-classical correspondence stands only for high quantum numbers.  
%%%%%%%%%%%%%%%%%%%%%%%%%%%%%%

As far as we take $\iota = \sqrt{\eta } \theta (\eta )$ and
we consider the negative part of the spectrum
$\mathcal{E} = -\mid \mathcal{E}\mid$, the function
$\theta $ obeys the equation

\begin{eqnarray}
\label{twoeig2}
\frac{\delta ^2\theta }{\delta \eta ^2} +
\frac{1}{\eta }\frac{\delta \theta }{\delta \eta } +
\left(\mid \mathcal{E}^{\prime }\mid  -
\frac{q^2}{\eta ^2}\right)
\theta = 0 \\                     
\mathcal{E}^{\prime } \equiv
\hbar ck^2\mathcal{E}, \qquad
q^2 \equiv \frac{1}{4}\left(1 - 128p^2\right)
\, .
\end{eqnarray}

Thus we see that a negative part of the spectrum exists
in correspondence to the solution

\begin{equation}
\theta (\eta ,\; \mathcal{E},\; p) =
AJ_q(\sqrt{\mid \mathcal{E}^{\prime }\mid } \eta ) +
BJ_{-q}(\sqrt{\mid \mathcal{E}^{\prime }\mid }\eta )
\, ,
\label{bessel}
\end{equation}

where $J_{\pm q}$ denote the corresponding
Bessel functions, while $A$ and $B$ are two integration
constants. This solution remains valid only as far as
$\mid p\mid < 1/(8\sqrt{2})$.

To give a precise
physical meaning to this picture, the following four main points
have to be addressed.\\ 
i) The existence of a stable ground level of negative
energy has to be inferred or provided by additional conditions.
ii) The spatial gradients
of the dynamical variables 
and therefore the associated super-momentum constraints,
have to be included into the problem and 
treated in a consistent way.
iii) The physical nature of the limit
$\eta \rightarrow 0$ has to be clarified by a
physical characterization of the dynamics.
iv) In order to restore general covariance in the classical limit,
$\mathcal{E}$ has to vanish for $\hbar \rightarrow 0$.
But, to be retained in the zero-order WKB approximation
$\mathcal{E}$ should behave like $\hbar ^{1-b}$ ($b>0$).\\ 

We conclude this section by stressing that, in the quantum regime, the values available
for $\mathcal{E}$ are provided by the super-Hamiltonian
spectrum. Thus, they depend on the boundary conditions
fixed for the system, 
but not on the initial form of the wavefunctional.
As a consequence, the induced
(quantum) fluid is determined by the intrinsic properties
of the geometrodynamics and the test character of this dust 
is no longer ensured.

\section{EVOLUTIONARY QUANTUM COSMOLOGY}
In order to investigate the implications of the synchronous 
quantum dynamics, we now present results, based on the works \cite{KL63} and \cite{BM04}, on the behavior of a generic inhomogeneous Universe. In fact, the absence of
specific symmetries is required by the impossibility of preserving
them in quantum cosmology at super-horizon scales.\\ 
The quantum implementation into an evolutionary framework will be presented in the following subsection.

\subsection{GENERIC COSMOLOGICAL SOLUTION} 
As shown in \cite{BM07}, a generic inhomogeneous cosmological model (in terms of Misner variables $\alpha$ and $\beta_\pm$) is described by the action
\begin{equation}
\label{sred2}
S_{Red} = \int_{\Sigma ^3_t\times\mathcal{R}}dt d^3 y
\left\{ p_{\alpha }\partial_t \alpha
+ p_+\partial_t \beta _+
+ p_-\partial_t \beta _- 
+ p_{\varphi }\partial _t \varphi   
- H\right\}
\end{equation}
\begin{equation}
H = \frac{c^2ke^{-3\alpha }}{3}\left[ 
-p_{\alpha }^2 + p_+^2 + p_-^2 \right]
+ \frac{3}{8\pi } p_{\phi }^2
- U(\alpha ,\; \beta _{\pm })
+ \frac{\Lambda }{k}e^{3\alpha }
\end{equation}
\begin{equation}
U = \frac{1}{2k\mid J\mid ^2}e^{\alpha } V(\beta _{\pm }), \quad
V(\beta _{\pm}) =
\lambda _1^2e^{4\beta _+ + 4\sqrt{3}\beta _-} +
\lambda _2^2e^{4\beta _+ - 4\sqrt{3}\beta _-} +
\lambda _3^2e^{-8\beta _+} 
\, ,
\end{equation}

${y^a}$ being a suitable set of spatial coordinates, with $J$ the Jacobian of the transformation $y^a=y^a(x^l)$.
Here the functions $\lambda _a(y^b(x^l))$
fix the model inhomogeneity.

Above, to account for the inflationary scenario,
we included in the dynamics a massless
scalar field $\varphi$ and a cosmological constant $\Lambda $.
The presence of these two terms allow us to model the main features
of the inflaton field dynamics in the pre-inflation and
slow-rolling phases, respectively.
However, both $\varphi$ and $\Lambda$ have also an important
dynamical role. In fact, 
on one hand, the presence of this scalar field is
crucial to neglect the potential term $U$
on a classical \cite{B00,M00} and a quantum level \cite{K92}.
The kinetic term of $\varphi$ is able to destroy the
chaotic behavior induced by the spatial curvature, and therefore
here no serious differences are expected in comparison with
to the loop quantum gravity approach (in such a
formulation, the chaoticity would disappear even without the scalar field
\cite{B04}).
On the other hand, the later de-Sitter dynamics,
associated to the slow-rolling regime,
provides the isotropization of
the causal homogeneous portions of the Universe \cite{KM02},
justifying the estimations
(based on the actual Universe parameters) 
which we will address below for our generic model. 

\subsection{THE QUANTUM DYNAMICS}

Since the total Hamiltonian of the system reduces, near the
singularity, to the sum of $\infty ^3$ independent point-like
contributions, the Wheeler superspace is decomposed into
$\infty ^3$ minisuperspaces and the Schr\"odinger
functional equation splits correspondingly.
Fixing the space point $x^l$
({\it i.e.} $y^a(x^l)$), the quantum dynamics reads
(we denote by the subscript $x$ any minisuperspace quantity)

\begin{eqnarray}
\label{sch}
i\hbar \partial _t \psi _x = 
\hat{H}_x\psi _x =  \frac{c^2\hbar ^2k}{3} \left[ 
\partial _{\alpha }e^{-3\alpha }\partial _{\alpha}
-e^{-3\alpha }\left( \partial ^2_+ + \partial ^2_-  
\right) \right]\psi _x 
- \frac{3\hbar ^2}{8\pi } e^{-3\alpha }\partial ^2_{\varphi }\psi _x
- \nonumber\\ 
- \left( \frac{1}{2k\mid J\mid ^2}e^{\alpha } V(\beta _{\pm })
- \frac{\Lambda }{k}e^{3\alpha }\right)  \psi _x\\
\psi _x = \psi _x (t , \; \alpha , \; \beta _{\pm} , \; \varphi )
\, .
\end{eqnarray}

We now take the following integral representation for the
wave function $\psi _x$

\begin{eqnarray}
\label{exp}
\psi _x = \int 
d\mathcal{E}_x \mathcal{B}(\mathcal{E}_x)
\sigma _x(\alpha , \; \beta _{\pm } , \; \varphi ,\; \mathcal{E}_x)
exp \left\{ -\frac{i}{\hbar }\int _{t_0}^tN_x\mathcal{E}_xdt^{\prime }
\right\}\\
\sigma _x = \xi _x(\alpha , \; \mathcal{E}_x)
\Pi _x(\alpha , \; \beta _{\pm } , \; \varphi )
\, , 
\end{eqnarray}
where $\mathcal{B}$ is 
fixed by the initial conditions at $t_0$.
Hence, we get the following reduced problems

\begin{eqnarray}
\label{eigenvp}
\hat{H}\sigma _x = \mathcal{E}_x \sigma _x\\ 
\left( -\partial ^2_+ - \partial ^2_-
- \frac{9\hbar ^2}{8\pi c^2k } \partial ^2_{\varphi }
\right) \Pi _x
- \frac{3}{2c^2\hbar ^2k^2\mid J\mid ^2}e^{4\alpha } V(\beta _{\pm })\Pi _x =
v^2(\alpha ) \Pi _x\\ 
\left[ \frac{c^2\hbar ^2k}{3}\left( 
\partial _{\alpha }e^{-3\alpha }\partial _{\alpha } \xi _x
+ e^{-3\alpha }v^2(\alpha ) \right) +
\frac{\Lambda }{k}e^{3\alpha }\right] \xi _x
= \mathcal{E}_x\xi _x
\, . 
\end{eqnarray}

Above, in deriving the equation for $\xi _x$, we neglected the
dependence of $\pi _x$ on $\alpha $ because, asymptotically to
the singularity ($\alpha \rightarrow -\infty$),
it has to  be of higher
order ({\it i.e.} we address an adiabatic approximation).
If we take the plane wave solution
$\pi _x 
\propto e^{i(v_+\beta _+ + v_-\beta _- + v_{\varphi }\varphi )}$,
then we get

\begin{equation}
v^2 \equiv v_+^2 + v_-^2 
+ \frac{9}{8\pi c^2k}v_{\varphi }^2 = const.
\label{pw}
\end{equation}
and, in the limit $\alpha \rightarrow -\infty$, this choice is
a good approximation as far as the following condition holds

\begin{eqnarray}
\label{cond}
v^2_{\beta }\equiv v^2_+ + v^2_- \gg
\frac{3e^{4\alpha }}{2c^2\hbar ^2k^2\mid J\mid ^2}
\mid \bar{V}\mid,\qquad
\bar{V} \equiv \frac{1}{\Delta \beta }
\int _{\Delta \beta ^2}d^2\beta \left\{ 
V(\beta _{\pm })\right\}
\, .
\end{eqnarray}

Here, instead of ideal monochromatic solutions,
we considered wave packets which are flat over the
width $\Delta \beta \sim 1/\Delta v_{\beta }\gg 1$
($\Delta v_{\beta }$ being the standard deviation in the
momenta space). 

Once the new variable
$\tau = e^{3\alpha }$ is adopted,
the above eigenvalues
problem for the wavefunction $\xi _x$ reads

\begin{equation}
\frac{c^2\hbar ^2k}{3} \left( 
9\frac{d ^2 }{d \tau ^2}
+\frac{v^2}{\tau ^2}  
\right) \xi _x + 
\frac{\Lambda }{k}\xi _x
= \frac{\mathcal{E}_ x}{\tau }\xi _x
\, .  
\label{ph9}
\end{equation} 

Here the potential term reads
$\mathcal{O}\left( \tau ^{-2/3}\right)$
and therefore the cosmological term dominates
as far as the following condition holds

\begin{equation}
L_{\Lambda } \equiv \frac{1}{\sqrt{\Lambda }} \ll
L_{in} \equiv \frac{\tau ^{1/3}}{\langle \lambda \rangle }
\, ,
\label{coninf}
\end{equation}

where $\langle \lambda \rangle$ denotes an
average value on the functions $\lambda _a$.
The above relation corresponds to the request that the
length scale associated to the ``vacuum energy''
($L_{\Lambda }$) is much less then the physical scale of
the Universe inhomogeneity
($L_{in}$), that is just one of the usual constraints for starting the
inflation.

Searching for a solution to equation
(\ref{ph9}) in the form
$\xi _x = \tau ^{\delta }f_x(\tau )$, we get

\begin{eqnarray}
\label{solint}
\delta = \frac{1}{2}\left( 1 \pm
\sqrt{1 - \frac{4}{9}v^2}\right)\\
\frac{d^2f_x}{d\tau ^2} + 
\frac{1}{\tau }
\left( 2\delta \frac{df_x}{d\tau }
- \frac{1}{3L_{\mathcal{E}}l_P^2}\right) + 
\frac{1}{3L_{\Lambda }^2l_P^4}f_x = 0
\, ,  
\end{eqnarray} 

$L_{\mathcal{E}} = \frac{\hbar c}{\mathcal{E}}$ being the
characteristic length associated to the Universe ``energy'', 
while $l_P \equiv \sqrt{\hbar ck}$
denotes the Planck scale length.\\
Hence, it is easy to check that, within the precision of our
potential-free regime, the solution of the above equation 
admits an exponential form
(as boundary conditions, we require that $\xi $ vanishes
in correspondence to the singularity in $\tau = 0$ and
decays at large $\tau $, where the potential becomes relevant),
{\it i.e.}

\begin{eqnarray}
\label{ph10}
f = \mathcal{C}
exp \{ -\beta ^2\tau ^2 + \gamma \tau \} \\
\gamma = 2\mid \beta \mid
\sqrt{\delta + \frac{1}{2}                 
-\frac{1}{12L_{\Lambda }^2l_P^4\beta ^2}},\qquad
\frac{1}{L_{\mathcal{E}}l_P^2} =
6\delta \gamma 
\, . 
\end{eqnarray}

We see that the 
quantum dynamics in a
fixed space point
({\it i.e.} over a causal portion of the Universe)
is described, in the considered approximation
($\tau \ll 1$), by a free wavepacket for the variables
$\beta _{\pm }$ and $\varphi$  and by a profile in $\tau $
which has a maximum in
$\tau = (\gamma + \sqrt{\gamma ^2 + 8\delta \beta^2})/4\beta^2$.\\
We stress that above the constant $\mathcal{C}$ has to be
regarded in $\sigma _x$ as a generic function of the quantum numbers
$\{ k_{\pm }, \; k_{\varphi }\}$.

To estimate the range of variation for the
eigenvalue $\mathcal{E}_ x$,
{\it i.e.} the length $L_{\mathcal{E}}$,
we observe that
the maximum value admissible for the quantity
$\delta $ corresponds to $v^2=0$, {\it i.e.}
$\delta=1$. Furthermore, the validity
of the solution above requires that the
condition
$\beta ^2\tau \ll \gamma =
2\sqrt{\delta + \frac{1}{2}                 
-\frac{1}{12L_{\Lambda }^2l_P^4\beta ^2}}
\mid \beta \mid$
(which implies
$\mid \beta \mid \tau \ll
\sqrt{6}
\sqrt{1 
-\frac{1}{18L_{\Lambda }^2l_P^4\beta ^2}}$) holds.\\

In agreement with the idea that the gravitational
field has a natural lattice structure on the Planckian
scale \cite{RS03}, we can take as minimal value for the
variable $\tau $, the amount $l_{Pl}^3$.
Putting together these considerations, we arrive to
the fundamental inequality 

\begin{equation}
\mid \beta \mid \ll
\frac{\sqrt{6}}{l_P^3}
\sqrt{1 
-\frac{1}{18L_{\Lambda }^2l_P^4\beta ^2}}
\label{fininequa}
\, .
\end{equation}

The reality of the square root
($\delta $ and $\beta $ have to be real to ensure the
reality of $\mathcal{E}$) requires that
$\mid \beta \mid \ge \frac{1}{3\sqrt{2}L_{\Lambda }l_P^2}$.
If, as expected, $L_{\Lambda }\gg l_P$, then
the above inequality (\ref{fininequa}) reads
$\mid \beta \mid \ll 1/l_P^3$
(because the neglected term behaves as
$\mathcal{O}\left(l_P^2/L_{\Lambda }^2\right) \ll 1$),
or equivalently

\begin{equation} 
\mid \mathcal{E}_ x\mid \ll \frac{c^2k \hbar ^2}
{l_{Pl}^3} \sim \mathcal{O}(M_{Pl}c^2)
\rightarrow L_{\mathcal{E}} \gg l_P
\, ,
\label{ph11}
\end{equation} 

where $M_{Pl} \equiv \hbar /(l_{Pl}c)$ is the Planck mass.\\
It is worth noting that here the appearance of a stable
ground state for the model is a consequence of
the cut-off request. According to the standard
interpretation of quantum mechanics,
we make the assumption that the Universe must necessarily
approach this state of minimal ``energy'' as a result of its
spontaneous evolution.

As shown in \cite{M03}, the above range of variation for
the super-Hamiltonian eigenvalue leads,
in the isotropic case, 
to a negligible contribution of this term toward the actual
Universe critical parameter, when an inflationary stage
is (like here) addressed. In fact, to estimate the critical
parameter associated to the new contribution, say
$\Omega _{\mathcal{E}}$, we observe that the super-Hamiltonian
eigenvalue, in the classical limit, behaves as a constant
of motion and therefore it provides today the energy density
$\rho _{\mathcal{E}} \ll \mathcal{O}((M_{Pl}c^2)/R_0^3)$
($R_0 \sim \mathcal{O}(10^{28}cm)$ denoting the present Universe
radius of curvature). Since the actual critical density can be
expressed as
$\rho _c \sim \mathcal{O}(c^4/[GR_0^2(\Omega - 1)])$
(being $\Omega = 1 \pm \mathcal{O}(10^{-2})$ the total Universe
critical parameter), then we have

\begin{equation}
\Omega _{\mathcal{E}} \equiv
\frac{\rho _{\mathcal{E}}}{\rho _c} \ll
\mathcal{O}\left( \frac{10^{-2}GM_{Pl}}{c^2R_0} \right) \sim 
\mathcal{O}\left( \frac{10^{-2}l_{Pl}}{R_0} \right) \sim 
\mathcal{O}\left( 10^{-60}\right)
\, .
\label{crit}
\end{equation}

Thus, to regard the dust fluid as a \emph{dark matter candidate}, ``matter'' (from the thermal bath) 
must play a relevant role in the
Planckian Universe evolution (see the model addressed in
\cite{CM04,BM06}, where ultrarelativistic matter
and a perfect gas were included).
The correspondence between the isotropic case and
the generic dynamics is possible because the last is
homogeneous at the horizon scale, and the anisotropies contribution
is isomorphic to the scalar field one
(both providing a free energy density
$\propto e^{-6\alpha }$).

\section{THE QUASI-CLASSICAL LIMIT}

Though we solved equation (\ref{ph9}) only in
the limit of small values of $\tau $,
where the spatial curvature is negligible, 
nevertheless we now show that conditions  
for the classical limit of the $\tau$-dynamics exist
within such approximation. For a discussion on the decoherence 
of the scale factor in a Freedam-Robertson-Walker space-time see \cite{K87} (for the semi-classical limit of the Wheeler-DeWitt dynamics in a more general case see \cite{KS91}).

In the variables
$\{ \tau ,\; \beta _{\pm}, \; \varphi \}$, 
the local minisuperspace line element reads

\begin{equation}
d\Gamma ^2 = -\frac{1}{3c^2k\tau }d\tau ^2 + 
\frac{3\tau }{c^2k }\left( d\beta _+^2 +
d\beta _-^2 \right) + \frac{8\pi \tau }{3}d\varphi ^2
\label{msm}
\, .
\end{equation}

Thus, for small Universe volumes, 
we construct the semiclassical limit
of the considered model, applying
a procedure in the spirit fixed in \cite{V89} and allowed
by the decoupling of the asymptotic classical $\tau$-dynamics
from all other variables.\\
Aim of the analysis here faced is
to separate the evolution 
of the quasi-classical variable $\tau$, 
from the quantum anisotropies
$\beta _{\pm}$ and the scalar field $\varphi$, for which an Hilbert
space can then be defined.\\
Having in mind this picture and fixing,
by the line element above, the timelike
variable $\tau $ as the quasi-classical component of the Universe,
we can take the (point-like)
wave function $\sigma $ in the form

\begin{equation}
\sigma = \mu (\tau )
exp\left\{ i\frac{\Phi (\tau )}{\hbar }\right\}
\mathcal{P}(\tau, \; \beta _{\pm }, \; \varphi )
\label{scwf}
\, .
\end{equation}

Substituting this expression into the eigenvalue
problem (\ref{eigenvp}),
taking the limit $\hbar \rightarrow 0$
({\it i.e.} $\tau \gg l_P^3$) and addressing the approximations fixed in
\cite{V89}, then we get the following system of
three coupled equations

\begin{eqnarray}
\label{ph13}
-3c^2k  
\left( \frac{ d\Phi }{d\tau }\right) ^2 +
\frac{\Lambda }{k}
- \frac{\mathcal{E}}{\tau } = 0 \\
\frac{d }{d \tau }
\left( \mu ^2 \frac{d \Phi }{d \tau }
\right) = 0\\ 
6i\hbar c^2k\frac{d \Phi }{d\tau }
\frac{\partial \mathcal{P}}{\partial \tau } = 
 \frac{1}{\tau ^2}\left[ \frac{\hbar ^2c^2k}{3}\left(\partial ^2_+
+ \partial ^2_- \right) 
+ \frac{3}{8\pi } \partial ^2_{\varphi }
\right] \mathcal{P} 
\, .
\end{eqnarray} 

The first equation gives the Hamilton-Jacobi dynamics
and therefore the identification $p_{\tau } = d\Phi /d\tau$ naturally
arises ($p_{\tau }$ being the conjugate momentum to the variable
$\tau $).
Starting from the action (\ref{sred2}),
it is easy to check the relation
$-6c^2kp_{\tau } = \partial _t\tau /\tau$. Hence the equation for the
wave function $\mathcal{P}$ takes the meaningful Schr\"odinger form 

\begin{equation}
i\hbar \partial _t\mathcal{P} = 
-\frac{1}{\tau (t)}\left[ \frac{\hbar ^2c^2k}{3}\left(\partial ^2_+
+ \partial ^2_- \right) 
+ \frac{3}{8\pi } \partial ^2_{\varphi }
\right] \mathcal{P} 
\label{redsch}
\, ,
\end{equation}

where the function $\tau (t)$ is assigned by the classical dynamics
(containing the dust term too).
In the present evolutionary scheme it comes out from the equation

\begin{equation}
t = \frac{\partial \Phi }{\partial \mathcal{E}} + t^* , \;
t^* = const.
\, .
\end{equation}

Since the Universe lies expectantly in the ground state
of negative energy ({\it i.e.} we take
$\mathcal{E} = - \mid \mathcal{E}\mid$),
then, in the region
$\tau \ll L_{\mathcal{E}}L_{\Lambda }^2$,
the Hamilton-Jacobi equation and the coupled one for
the amplitude $\mu $, admit the solutions

\begin{eqnarray}
\label{ph14}
\Phi = 2\sqrt{\frac{\mid \mathcal{E}\mid}{3c^2k}\tau}\qquad
\mu = \mathcal{D}\sqrt[4]{\tau }\qquad 
\mathcal{D} = \mathcal{D}(x^i) 
\, ;
\end{eqnarray}

above, we set to zero the integration
constant for $\Phi$,
because it provides only a phase factor.\\
It can be verified that such behaviors of
$\Phi$ and $\mu$ are in agreement with the
approximations at the ground of the system (\ref{ph13}).\\
The solution of the Schr\"odinger equation
(\ref{redsch}) has exactly the same form of a 3-dimensional
non-relativistic free particle, as soon as the time variable
$\mathcal{T} = \int(dt/\tau(t))$ is adopted.

Thus we show how the energy spectrum
arises near the cosmological singularity.
Indeed, the question concerning the classical limit of
such degrees of freedom remains open. The description of a transition,
from the quantum mixmaster to a classical isotropic
Universe, remains an open topic in theoretical
cosmology and the different proposals for its
solution strictly depend on the initial condition
on the system (for a discussion in the Wheeler-DeWitt
approach see \cite{KM97,BBC04}). In the present context, we
stress that, sooner or later (out of our approximation
scheme), the potential terms, both for the anisotropic
variable and the scalar field, would become important
in the evolution. It is just in this phase that we
expect the appearance of a classical behavior.
This point of view is supported by the quadratic
feature that such potential terms must approach.
The potential of the mixmaster becomes quadratic
in $\beta _{\pm}$ as far as the Universe expansion
({\it i.e.} increasing values of $\tau$)
frozes out the corresponding anisotropy
({\it i.e.} small values of $\beta _{\pm}$ are approached).
The potential term associated to the scalar field
is quadratic near its stable minimum, which
must exist before the spontaneous symmetry breaking
of the inflationary scenario. Therefore, wave-packets peaked around small values of $\beta_\pm$ and $\varphi$ seem favorable in reaching the classical limit.\\
When the system falls into this quadratic approximation,
stable coherent states can be constructed with
classical properties.

\section{FROM THE SYNCHRONOUS REFERENCE TO A
GENERAL POINT OF VIEW}

The results of the above discussion outline
that synchronous quantum gravity is an evolutionary theory
of the space-time, but the associated phenomenology  seems to be
compatible, on a cosmological level, with the Wheeler-DeWitt
paradigm: both the approaches provide General Relativity in the
classical limit for $\hbar \rightarrow 0$. The key feature here is
the dependence on $\hbar$ acquired by the super-Hamiltonian eigenvalue. This picture suggests us to
investigate for more general contexts which would
predict evolving wave-functionals only up to
some order in $\hbar$. In this respect, we fix
our attention on the relations existing between
a statistical representation of a stochastic
system and its semi-classical features as coming
out of the WKB limit. 

The physical reason leading us to compare
the semi-classical limit of the quantum mechanics
to an ensamble picture is that, for a stochastic
system, two independent (one classical and another
semi-classical) probability distributions make sense.
Indeed, for any classical dynamics we could define
a probability distribution as a delta functional
over the deterministic trajectory. Despite this
choice would naturally imply the necessity of
an evolutionary approach, nevertheless it appears
rather ill-defined to be properly addressed
(for a discussion of this point of view, as well as
of a Bohmian approach to the same question see
\cite{Ni03,K95-97}).

This parallelism fixes, for stochastic gravitational systems, a correspondence between the time evolution of the ensemble distribution and that of the first correction in $\hbar$ to the wave-functional.

\subsection{THE GRAVITATIONAL FIELD}

Let us now analyze the case of the gravitational
field, with the aim of inferring an appropriate equation
for its quantum dynamics.

In a generic
reference frame, the action describing
the gravitational field reads as follows
\cite{ADM62} 

\begin{equation}   
S = \int _{\mathcal{V}4} \left\{
\pi ^{ij}\partial _th_{ij}
- NH - N^iH_i \right\} d^3xdt
\label{ag1} 
\end{equation}  

where (here we restate in detail the notation)
$\mathcal{V}^4 = \Sigma ^3_t\times\mathcal{R}$ denotes
the whole 4-manifold
(sliced into the 1-parameter family of
compact boundaryless 3-hypersurfaces $\Sigma ^3_t$),
$\pi ^{ij}$ are the
conjugate momenta to the   
3-dimensional metric tensor $h_{ij}$, 
while the gravitational
super-Hamiltonian 
$H$ and the supermomentum $H_i$ take 
the form

\begin{equation}   
\label{ag2}
H \equiv \frac{16\pi G}{c^2}G_{ijkl}\pi^{ij}\pi^{kl}
- \frac{c^4}{16\pi G}
\sqrt{h}{}^3R,  \, \quad G_{ijkl} \equiv \frac{1}{2\sqrt{h}}
(h_{ik}h_{jl} + h_{il}h_{jk} - h_{ij}h_{k})
\end{equation} 
\begin{equation}
H_i \equiv -2~^3\nabla _j\pi ^j_i 
\, .
\end{equation}

In the above expressions, ${}^3R$
and $~^3\nabla_i(\; )$
denote 
the Ricci scalar and the covariant
derivative constructed
by the 3-metric $h_{ij}$ respectively, while 
$h \equiv deth_{ij}$.

The dynamics of the system is summarized
by the following field equations
(obtained variating the action  
with respect to $N$, $N^i$, $\pi^{ij}$ and $h_{ij}$)

\begin{eqnarray}
\label{ag3}
H = 0 \; , \; H_i = 0 \\
\partial _th_{ij} =
\frac{\delta     \mathcal{H}}{\delta \pi ^{ij}}
\; , \;
\partial _t\pi ^{ij} =
-\frac{\delta \mathcal{H}}{\delta h _{ij}} \\
\mathcal{H}\equiv \int_{\Sigma ^3_t}d^3x
\,  
\left\{ NH + N^iH_i\right\}.
\end{eqnarray}

The four constraints $H = H_i = 0$
reflect the 4-diffeomorphism invariance
of General Relativity and they 
are characterized by the following two
properties:

\vspace{0.6cm}

i) These constraints are non-evolutionary,
{\it i.e.} if they are satisfied
by the Cauchy data
on the initial hypersurface
(say at $t = t_0$),
then they remain valid for all the evolution,
in view of the Hamilton equations.

ii) The four constraints lead,
under the replacement
$\pi ^{ij} = \delta S/\delta h_{ij}$, 
to the Hamilton-Jacobi equations 

\begin{eqnarray}   
\label{ag4}
\widehat{HJ}S        
\equiv \frac{16\pi G}{c^2}G_{ijkl}
\frac{\delta S}{\delta h_{ij}}
\frac{\delta S}{\delta h_{kl}}
- \frac{c^4}{16\pi G}
\sqrt{h}{}^3R = 0 \\
\widehat{HJ}_iS \equiv
-2h_{il}~^3\nabla _j
\frac{\delta S}{\delta h_{jl}} = 0
\, .
\end{eqnarray}  

This set of equations provides alone the
whole gravitational field dynamics.

\vspace{0.6cm}

Thus, if we restrict the phase space
of a stochastic gravitational system 
to ensembles
which satisfy the constraints above
({\it i.e.} to the Wheeler phase superspace),
then the Hamilton
equations fix the dynamics of the system, for which the
lapse function $N$ and the shift vector $N^i$ play the
role of parametric functions.
In analogy to the non-relativistic particle, 
the continuity equation for the ensemble density
(functional)
$\varepsilon (t,\; N(t,\; x^l), \; N^i(t,\; x^l),\;
h_{ij}(x^l), \pi ^{ij})(x^l))$
reads as

\begin{equation}
\partial _t\varepsilon + 
\int _{\Sigma ^3_t}d^3x\frac{\delta }{\delta h_{ij}}
\left\{
\frac{\delta \mathcal{H}}{\delta  \pi ^{ij}}    
\varepsilon \right\}
-\int _{\Sigma ^3_t}d^3x\frac{\delta }{\delta \pi ^{ij}}
\left\{
\frac{\delta \mathcal{H}}{\delta  h _{ij}}  
\varepsilon \right\} = 0
\, . 
\label{ag5}
\end{equation}

Making use of the following relation 

\begin{equation}
\frac{\delta \mathcal{H}}{\delta  \pi ^{ij}}  = 
\frac{32\pi G}{c^2}G_{ijkl}\pi^{kl}
+ ~^3\nabla _iN_j +~^3\nabla _jN_i \\
\, , 
\label{ag6}
\end{equation}
restricting the phase space distribution to the form
$\varepsilon = \varrho (t, \; h_{ij})\delta
\left( \pi ^{ij} - \delta S/\delta h_{ij}\right)$, 
and evaluating the integral over the momentum space,
we arrive to the reduced continuity equation

\begin{equation}
\partial _t\varrho
+ \frac{32\pi G}{c^2}
\int _{\Sigma ^3_t}d^3x\frac{\delta }{\delta h_{ij}}
\left\{
G_{ijkl}\frac{\delta S}{\delta h_{kl}}
\varrho \right\}
+ 2\int _{\Sigma ^3_t}d^3x\frac{\delta }{\delta h_{ij}}
\left\{
~^3\nabla _iN_j
\varrho \right\} = 0 
\, , 
\label{ag7}
\end{equation}
$\varrho \equiv \int \varepsilon D\pi $
being the distribution reduced to the configuration
space (here $D\pi $ denotes the Lebesgue measure defined for the conjugate
momentum).

Observing that (with obvious notation)

\begin{equation}
\frac{\delta }{\delta h_{ij}} \left(
~^3\nabla _iN_j\right)
= -\frac{1}{2}~^3\nabla _iN^i 
\label{Gamma}
\end{equation}
and since the hypersurfaces $\Sigma ^3_t$
are taken to be compact ones without boundary
(which allows us to eliminate
total divergences)
\footnote{We stress that the momentum $\pi ^{ij}$, as
well as $\delta \varrho /\delta h_{ij}$ are 3-tensor
densities of weight $1/2$.}, 
then the above equation (\ref{ag7}) rewrites

\begin{equation}
\partial _t\varrho
+ \frac{32\pi G}{c^2}
\int _{\Sigma ^3_t}d^3x\frac{\delta }{\delta h_{ij}}
\left\{
G_{ijkl}\frac{\delta S}{\delta h_{kl}}
\varrho \right\}
- 2\int _{\Sigma ^3_t}d^3x 
\left\{
N_j~^3\nabla _i
\frac{\delta \varrho }{\delta h_{ij}}
\right\} = 0 
\, . 
\label{ag7f}
\end{equation}

The dynamics of the system has to be invariant under
the infinitesimal 3-diffeomorphism
$x^{l^{\prime }} = x^l + \xi ^l (x^j)$
($\xi ^l$ being generic displacements), 
which induces 3-metric transformations
$h^{\prime }_{ij} = h_{ij} - 
2~^3\nabla _{(i}\xi _{j)}$.\\
Requiring that $\varrho $ is invariant
under such 3-metric changes, yields
$\varrho (h_{ij} - 2~^3\nabla _{(i}\xi _{j)}) - 
\varrho (h_{ij}) = \delta \varrho = 0$, {\it i.e.} 

\begin{equation}
\delta \varrho = -2\int _{\Sigma ^3_t}d^3x 
\left\{
\frac{\delta \varrho }{\delta h_{ij}}
~^3\nabla _i\xi _j
\right\} =
2\int _{\Sigma ^3_t} d^3x 
\left\{
~^3\nabla _i\left( \frac{\delta \varrho }{\delta h_{ij}}
\right) \xi _j\right\} = 0 
\, .
\label{ag8}
\end{equation}

Since $\xi ^i$ are generic space displacements, we get
$2~^3\nabla _j\left( \frac{\delta \varrho }{\delta h_{ij}} \right) = 0$.\\
In view of this result,
the ensemble density loses its
parametric dependence on the
shift vector and 
it takes values on the 3-geometries
$\{ h_{ij}\}$. As far as we average the constraint
$H = 0$ over the momentum space, we recover the
Hamilton-Jacobi equation and then 
the statistical properties
of the gravitational system
(as viewed in the Wheeler superspace) 
are summarized
by the following functional equations 

\begin{eqnarray}   
\label{ag9}
\frac{16\pi G}{c^2}G_{ijkl}
\frac{\delta S}{\delta h_{ij}}
\frac{\delta S}{\delta h_{kl}}
- \frac{c^4}{16\pi G}
\sqrt{h}{}^3R = 0 \\
\partial _t\varrho
+ \frac{32\pi G}{c^2}
\int _{\Sigma ^3_t}d^3x\frac{\delta }{\delta h_{ij}}
\left\{
G_{ijkl}\frac{\delta S}{\delta h_{kl}}
\varrho \right\} = 0 \\ 
\widehat{HJ}_iS = \widehat{HJ}_i\varrho = 0
\, . 
\end{eqnarray}  

If we develop the time dependence of $\varrho$
in Fourier series, {\it i.e.}

\begin{equation}
\varrho (t,\; \{ h_{ij}\} ) = \int_{-\infty }^{\infty }
d\omega \bar{\varrho } (\omega ,\; \{ h_{ij}\} )
e^{i\omega t}
\, ,
\label{ftg}
\end{equation}

then the second of the above equation rewrites as

\begin{eqnarray}
\frac{32\pi G}{c^2}
\int _{\Sigma ^3_t}d^3x\frac{\delta }{\delta h_{ij}}
\left\{
G_{ijkl}\frac{\delta S}{\delta h_{kl}}
\bar{\varrho }\right\} = -i\hbar \omega 
\bar{\varrho}   
\, . 
\end{eqnarray}  

This equation, together with the Hamilton-Jacobi
system and the condition for 3-diffeomorphisms
invariance, provides the statistical framework to be
used when fixing the WKB limit of the quantum dynamics.

In analogy to what done for the non-relativistic
particle, 
let us consider the following smeared eigenvalue
problem

\begin{eqnarray}
\label{ag10}
\left\{
\int_{\Sigma ^3_t}d^3xN\hat{H} \right\} \Psi
= E^Q\Psi \\
\hat{H}_i\Psi = 0\\
\Psi = \Psi (t,\; N,\; \{ h_{ij}\} )
\, ,
\end{eqnarray}

where the operators $\hat{H}$ and $\hat{H}_i$
are casted via the conjugate ones
$\hat{h}_{ij}$ and
$\hat{\pi }^{ij} = -i\hbar \delta (\quad )/\delta h_{ij}$
(for the sake of simplicity, here we take $l_P = 1$).
To safe the Hermitianity of the 
super-Hamiltonian, we are lead to
take the normal ordering (see \cite{M02}) 

\begin{equation}
G_{ijkl}\pi ^{ij}\pi ^{kl} \rightarrow
-\hbar ^2\frac{\delta }{\delta h_{ij}}
G_{ijkl}
\frac{\delta }{\delta h_{kl}}
\label{ag11a}
\end{equation}

Taking the expansion

\begin{equation}
\Psi = e^{\frac{i}{\hbar}\Sigma}
\,, \quad
\Sigma = \Sigma _0 + \frac{\hbar}{i}\Sigma _1
+ \left( \frac{\hbar}{i}\right) ^2
\Sigma _2 + ...
\, ,
\label{stfexc}
\end{equation}

then,
in the considered WKB limit and up to first
order in $\hbar$, from
(\ref{ag10}) we get the key relation

\begin{eqnarray}
\label{ag10xxcc}
\int_{\Sigma ^3_t}d^3x
\left\{ 
N\widehat{HJ}\Sigma _0 -i\hbar
\frac{\delta }{\delta h_{ij}}
\left(
G_{ijkl}\frac{\delta \Sigma _0}{\delta h_{kl}}
e^{2\Sigma _1}\right)e^{-2\Sigma_1} + \mathcal{O}\left( \hbar ^2\right) 
\right\} \Psi = \nonumber\\
= \left( E^Q_0 - E^Q_1
+ \mathcal{O}\left( \hbar ^2\right) \right) \Psi \\
\int _{\Sigma ^3_t}d^3x
\left\{ 
\widehat{HJ}_i\Sigma _0 - i\hbar
\widehat{HJ}_i\Sigma _1 \right\} = 0 
\, ,
\end{eqnarray}

The correspondence between this scheme and the ensemble
picture leads to the identifications
$S\equiv \Sigma _0$, $\varrho \equiv e^{2\Sigma _1}$
and $E^Q_0 = 0$, $E^Q_1 = \hbar \omega$. Thus, we see
that for a stochastic gravitational field, the
non-stationary character of the ensemble distribution
reflects the existence of a non-zero super-Hamiltonian
eigenvalue of order $\hbar$. This result is equivalent
to dealing with a Schr\"odinger equation (like in
Section 3), whose associated time evolution entirely
lives in the quantum sector, so ensuring the right classical
limit of General Relativity.

\subsection{THE EXAMPLE OF THE INHOMOGENEOUS MIXMASTER MODEL}

As an implementation of the above scheme, we now discuss the
asymptotic dynamics of the inhomogeneous mixmaster model
which is a widely-known example of a stochastic gravitational
system \cite{MTW73}. Here we do not address real new results, but we implement the well-established mixmaster picture to the present analogy between quantum and statistical geometrodynamics.\\ 
To get stochasticity we have to remove the
presence of a massless scalar field from the evolution of a generic Universe
toward the singularity (the cosmological term plays no role
asymptotically). Replacing the Misner variables
$\{ \alpha \; , \beta _+ \; ,\beta _-\}$ with the Misner-Chitr\`e-like
ones $\{ \rho \; , u\; ,v\}$, where $u$ and $v$
define the Poincar\`e half-plane representation of the
two-dimensional Lobachevsky space (for details of the coordinates
transformation see \cite{BM04}), the action
(\ref{sred2}) rewrites as 

\begin{equation}
\label{uvact}
S_{Red} = \int_{\Gamma_{Q}} d^{3} y
d\eta \left[p_{u} \frac{\partial u}{\partial \eta}
+ p_{v} \frac{\partial v}{\partial \eta}
+ p_{\rho } \frac{\partial \rho }{\partial \eta}
- \frac{N e^{-2\rho}}{24 D\mid J\mid } H  \right]
\end{equation}

with 

\begin{equation}
\nonumber 
H = -p_{\rho}^{2} + v^{2} \left(p_{u}^{2} + p_{v}^{2}
+ U(\rho \; , u\; .u)\right)
\end{equation}

and 

\begin{equation}
\nonumber 
D = \exp[-\sqrt{3} \frac{1+u+u^2+v^2}{v}  e^{\rho}].
\end{equation}

The potential term $U(\rho \; , u\; ,v)$ can be easily
calculated (see \cite{BM04,MTW73}). 

According to the analysis developed in the
previous subsection,
the ensemble representation of this stochastic system
takes the following form in the configuration space
associated to a space point
(in what follows we will omit the subscript
$y$ concerning the the point-like distribution
$w(\rho \; , u\; , v)$)

\begin{equation}
\label{nens}
-\left( \frac{\partial S}{\partial \rho}\right) ^2 +
v^2\left[ 
\left( \frac{\partial S}{\partial u}\right) ^2 +
\left( \frac{\partial S}{\partial v}\right) ^2
\right] + U(\rho \; , u\; .u) = 0
\end{equation}
\begin{eqnarray}
\frac{\partial w}{\partial t}
- \frac{N}{12\mid J\mid }
\frac{\partial \quad }{\partial \rho }\left(
e^{-2\rho }
\frac{\partial S}{\partial \rho }
\frac{w}{D}\right)+\nonumber\\ +
\frac{N}{12\mid J\mid }
e^{-2\rho }
\left[
v^2\frac{\partial \quad }{\partial u }\left(
\frac{\partial S}{\partial u }
\frac{w}{D}\right) +
\frac{\partial \quad }{\partial v }\left(
\frac{\partial S}{\partial v } v^2
\frac{w}{D}\right)
\right] = 0,
\end{eqnarray}

$S$ being the Hamilton-Jacobi function.

From the equation above, it can be easily inferred
that the limit toward the cosmological singularity 
$\rho \rightarrow \infty$
(where $D$ and all its derivatives vanish)
corresponds to asymptotically
increasingly smaller values of the
time derivative $\frac{\partial w}{\partial t}$.
This consideration holds only for a regular enough behavior
of the lapse function and it qualitatively
confirms that the ensemble distribution has to retain
a time dependence which, despite its low-order
character, accounts for the relic of an evolutionary
quantum gravity.\\
However, the correct characterization of the obtained
ensemble dynamics passes trough a careful discussion of
the allowance in fixing the lapse function. In fact,
due to the long-wavelength approximation, the spatial
gradients of the variable $\rho$ are asymptotically
negligible, so that it de-parametrizes
(in the line of \cite{BK95}) 
and the request

\begin{equation}
\partial _t\rho =
- \frac{Ne^{-2\tau }}{12D\mid J\mid}
\frac{\partial S}{\partial \tau} = 1 
\label{gaugg}
\end{equation}

can be imposed to deal with a real time coordinate.
When $\tau$ plays this role, the corresponding
ensemble picture is summarized by the
following system

\begin{eqnarray}
\label{nens1}
-\left( \frac{\partial S}{\partial \rho}\right) ^2 +
v^2\left[ 
\left( \frac{\partial S}{\partial u}\right) ^2 +
\left( \frac{\partial S}{\partial v}\right) ^2
\right] + U(\rho \; , u\; .u)
\equiv\nonumber\\\equiv
-\left( \frac{\partial S}{\partial \rho}\right) ^2 +
\left( \nabla S\right) ^2
+ U(\rho \; , u\; .u) = 0\\
\frac{\partial w}{\partial \rho} +
v^2\frac{\partial \quad }{\partial u }\left(
\frac{
\frac{\partial S}{\partial u }
}{\sqrt{ (\nabla S)^2 + U}}
w\right) +
\frac{\partial \quad }{\partial v }\left(
\frac{
v^2\frac{\partial S}{\partial v }
}{\sqrt{ (\nabla S)^2 + U}}
w\right) = 0.
\end{eqnarray}

When the asymptotic limit
$\{ \rho \rightarrow \infty ,\; U\rightarrow 0, \;
\frac{\partial S}{\partial \rho} = const.\}$
is taken, these equations overlap the stationary
picture described in some detail in \cite{BM07}
(see also references therein). However, as shown in
\cite{M01} (by using at all equivalent variables),
the stationary microcanonical distribution is
approached by an exponential decay in the $\rho$-dependence.
Such a feature quantitatively defines the time behavior
of the ensemble as a lower order effect for a point-like
mixmaster Universe.\\
But, the de-parametrization
of the variable $\rho$ and its time role in the dynamics
are consistent with a decoherence behavior as discussed
in the semiclassical limit above
(the main point here is that the asymptotic classical
evolution of $\rho$ is independent of the other variables).
Therefore, the correspondence between the evolution of the
microcanonical ensemble of the mixmaster and a
Schr\"odinger quantum gravity is valid in the 
limit when only some variables ($u$ and $v$ here)
follow a full
quantum behavior, while another one
($\rho$ here) is mainly a classical degree of freedom
(like in Section 5).
Of course, the possibility to deal with a component
of a gravitational system as a good time variable,
is not a general feature and the full correspondence
we established in this section would hold just for
those ensembles which make no allowance for any
decoherence scenario.\\
Finally, it is worth noting that the outlined picture
of the mixmaster chaoticity qualitatively
coincides with the one proposed in
\cite{PW83}, with respect to the definition of
an ensemble. In fact, the conclusion of
our analysis indicates that, in this model,
the chaoticity can be properly addressed by means
of a relational point of view.

\section{CONCLUDING REMARKS}

We proposed an evolutionary paradigm for the reformulation of 
the quantum gravity problem, based both on the restriction of the
covariance principle within a synchronous reference frame,
and on the more general correspondence between the ensemble dynamics
of stochastic gravitational systems and the semiclassical WKB limit
of their quantum dynamics. The common issue of these two different
approaches concerned the appearance of a non-zero eigenvalue of
the super-Hamiltonian, which turned out to vanish as
$\hbar \rightarrow 0$. Such contribution accounts for a
time evolution of the quantum gravitational field, but it
does not affect the right classical limit of General Relativity.

Dealing with the synchronous gauge, we get a non-zero super-Hamiltonian
eigenvalue following the scheme of the Noether theorem as applied
to the corresponding gravitational Lagrangian. The crucial point
here is that this additional term can be re-casted as a dust contribution,
which behaves as a source of the gravitational field. Thus,
we saw that the gauge fixing induces the appearance of a
real matter, playing the role of a reference. The quantum analysis
clarifies that such additional contribution has a non-classical
origin and, therefore, the limit of General Relativity is
always preserved as $\hbar \rightarrow 0$, even if we restricted
the dynamics to a synchronous reference (this point was discussed
in detail for the quantum cosmology model addressed above).

The merit of the discussion concerning the stochastic gravitational system, 
consists in the demonstration that
the ensemble time evolution would be associated with 
the first order in $\hbar$ in the expansion of the
super-Hamiltonian eigenvalue. 

It is worth stressing that the example of the chaotic inhomogeneous
mixmaster has outlined the necessity to deal with a Schr\"odinger
equation only in the decoherent picture, when a portion of
the system de-parametrizes from the whole and it plays the role
of a good time variable.

\ack
We thank Alessandra Corsi for her help in upgrading the English of the manuscript.


\section*{References}
\begin{thebibliography}{99}

\bibitem{D67} 
B. S. DeWitt, 
{\it Phys. Rev. }, (1967), {\bf 160}, 1113. 

\bibitem{ADM62}
R. Arnowitt, S. Deser, and C.W. Misner,
n Gravitation:
\textit{An Introduction to Current Research}, edited by L.Witten, Wiley, New
York, (1962) ,P.227.

\bibitem{K81}
K. V. Kuchar, 
in {\it Quantum Gravity II, a second Oxford symposium}, 
(1981), eds C. J. Isham et al., 
Clarendom Press., Oxford, 

\bibitem{I92}
C. J. Isham, 
{\it Canonical Quantum Gravity and the Problem of Time}, 
(1992), available /arxiv:/gr-qc/9201011.

\bibitem{SS04}
T. P. Shestakova and C. Simeone,
{\it Grav. Cosmol.}, (2004), {\bf 10}, 161.

\bibitem{Ba85}
T. Banks, {\it Nucl. Phys. B}, (1985), {\bf 249}, 332. 

\bibitem{Ro91a}
C. Rovelli, 
{\it Class. and Quantum. Grav. }, (1991), {\bf 8}, 297 and 317.

\bibitem{Ro91b} 
C. Rovelli, 
{\it Phys. Rev. D}, (1991), {\bf 43}, 442.

\bibitem{M02}
G. Montani, 
{\it Nucl. Phys. B}, (2002),
{\bf 634}, 370. 

\bibitem{MM04} 
S. Mercuri and G. Montani,
{\it Int. Journ. Mod. Phys. D}, (2004) {\bf 13}, 165.

\bibitem{KT91}
K. V. Kuchar and C. Torre,
{\it Phys. Rev. D}, (1991), {\bf 43}, 419. 

\bibitem{BK95}
J. D. Brown and K. V. Kuchar,
{\it Phys. Rev. D}, (1995), {\bf 51}, 5600.

\bibitem{H91a}
J. J. Halliwell and J. B. Hartle,
{\it Phys.Rev. D}, (1991), {\bf 43}, 1170.

\bibitem{H91b}
J. B. Hartle, in
{\em Conceptual Problems of Quantum Gravity},
(1991), edited by A. Ashtekar and J. Stachel
(Birkhauser, Boston).

\bibitem{H91c}
J. J. Halliwell,
{\it Phys. Rev. D}, (1991), {\bf 43}, 2590.

\bibitem{Tom}
S.Tomonaga, {\it Prog. Theor. Phys.},
(1946), {\bf 1}, 27; 
reprinted in ``Selected Papers on
Quantum Electrodynamics'',
edited by J. Schwinger (Dover, 1958). 

\bibitem{MM04b}
S. Mercuri and G. Montani,
{\it Mod. Phys. Lett. A}, (2004), {\bf 19}, n. 20, 1519.

\bibitem{KL63}
I M Khalatnikov and E M Lifshitz,
{\it Adv. Phys.}, (1963), {\bf 12}, 185.

\bibitem{BM04}
R. Benini and G. Montani, \emph{Phys. Rev. D},
(2004), {\bf 70}, 103527.

\bibitem{PW83}
D. N. Page, W. K. Wootters, \emph{Phys. Rev. D}, (1983), {\bf 27}, 2885.

\bibitem{IM01}
G. Imponente and G. Montani,
{\it Phys. Rev. D}, (2001), {\bf 63}, 103501.

\bibitem{M69}
C. W. Misner, (1969), {\bf 186}, n. 5, 1319.

\bibitem{B00}
B. Berger, {\it Phys. Rev. D}, (2000), {\bf 61}, 023508.

\bibitem{M00}
G. Montani, {\it Class. and Quantum Grav.}, (2000),
{\bf 17}, 2205.

\bibitem{K92}
A. A. Kirillov, \emph{JETP Lett.},
(1992), {\bf 55} 561.

\bibitem{B04}
M. Bojowald and G.Date,
{\it Phys. Rev. Lett}, (2004), {\bf 92}, 071302.

\bibitem{KM02}
A. A. Kirillov and G. Montani, {\it Phys. Rev. D}, (2002), {\bf 66}, 064010.

\bibitem{RS03}
C. Rovelli and S. Speziale, {\it Phys. Rev. D}, (2003), {\bf 67}, 064019.

\bibitem{M03}
G. Montani, 
{\it Int. Journ. Mod. Phys. D},
(2003), {\bf 12}, n. 8, 1445.

\bibitem{CM04}
G. Corvino and G. Montani,
{\it Mod. Phys. Lett. A}, (2004), {\bf 19}, n. 37, 2777.

\bibitem{BM06}
M. V. Battisti and G. Montani,
{\it Phys. Lett. B}, (2006), {\bf 637}, 203.

\bibitem{K87}
C. Kiefer, {\it Class. Quantum Grav.}, (1987), {\bf 4}, 1369.

\bibitem{KS91}
C. Kiefer, T. P. Singh, {\it Phys. Rev. D}, (1991), {\bf 44}, 1067.

\bibitem{M04}
G. Montani, 
{\it Int. Journ. Mod. Phys. D},
(2004), {\bf 13}, n.8, 1703.

\bibitem{V89}
A. Vilenkin, {Phys. Rev. D}, (1989), {\bf 39}, n. 4, 1116.


\bibitem{KM97}
A.A. Kirillov, G. Montani, {\it JETP Lett.}, {\bf 66}, (1997), 475-479.

\bibitem{BBC04}
B. Bolen, L. Bombelli, A. Corichi, {\it Class. Quantum Grav.}, {\bf 21}, (2004), 4087–4105.

\bibitem{K58} 
C. Kittel, {\it Elementary Statistical Physics}, 
(1958), edited by John Wiley et Sons, Inc..

\bibitem{Ni03}
H. Nikolic, arXiv:gr-qc/0312063.

\bibitem{K95-97}
S. P. Kim, {\it Phys. Rev. D}, (1995), {\bf 52}, 3382.\\
C. Bertoni, F. Finelli, and G. Venturi, {\it Class. Quantum Grav.}, (1996), {\bf 13}, 2375.\\
S. P. Kim, {\it Phys. Rev. D}, (1997), {\bf 55}, 7511.

\bibitem{MTW73}
C. W. Misner, K. Thorn, J. A. Wheeler,
\emph{Gravitation}, 
edited by W. H. Freeman et Company, 
New York, (1973), Chapter 21.

\bibitem{BM07}
R. Benini and G. Montani,
{\it Class. Quantum Grav.}, {\bf 24}, (2007), 387.

\bibitem{M01}
G. Montani,  {\it Nuovo Cimento}, {\bf 116 B},
(2001), 1375.

\end{thebibliography}
\end{document}